\documentclass[reqno,twoside,12pt]{amsart}
\makeatletter
\def\serieslogo@{}
\makeatother
\makeatletter
\def\@setcopyright{}
\makeatother

\newtheorem{thm}{Theorem}[section]
\newtheorem{cor}[thm]{Corollary}
\newtheorem{lem}[thm]{Lemma}

\theoremstyle{definition}

\theoremstyle{remark}

\begin{document}

\title{Newtonian limit of Maxwell fluid flows}
\author{Wen-An Yong}


\address{Zhou Pei-Yuan Center for Applied Mathematics\\
Tsinghua University\\
Beijing 100084, China}

\email{wayong@tsinghua.edu.cn} \keywords{Maxwell fluid flows,
hyperbolic relaxation systems, entropy dissipation, Newtonian limit,
error estimates}


\begin{abstract}
In this paper, we revise Maxwell's constitutive relation and
formulate a system of first-order partial differential equations
with two parameters for compressible viscoelastic fluid flows. The
system is shown to possess a nice conservation-dissipation
(relaxation) structure and therefore is symmetrizable hyperbolic.
Moreover, for smooth flows we rigorously verify that the revised
Maxwell's constitutive relations are compatible with Newton's law of
viscosity.
\end{abstract}

\maketitle \markboth{W.-A. Yong}{Newtonian limit of Maxwell fluid
flows}

\section{Introduction}

Maxwell fluids are among macromolecular or polymeric fluids. A large
number of experiments indicate that polymeric fluids exhibit elastic
as well as viscous properties \cite{BAH}. Thus, they are quite
different from small molecular fluids. The latter have viscosity as
the main feature, are satisfactorily characterized by
Newton's law of viscosity
\begin{equation}\label{11}
\tau = - \nu\Big[\nabla v + (\nabla v)^T -\frac{2}{3}\nabla \cdot v I\Big] - \kappa\nabla\cdot v I,
\end{equation}
and are also called Newtonian fluids. Here $\tau =\tau(x, t)$ is the
stress tensor of the fluid at space-time $(x, t)$, $\nu$ is the
shear viscosity, $\kappa$ is the bulk viscosity, $v=v(x, t)$ is the
velocity, $\nabla$ is the gradient operator with respect to the
space variable $x=(x_1, x_2, x_3)$, the superscript $T$ stands for
the transpose operator, and $I$ denotes the unit matrix of order 3.
Combining Newton's law of viscosity with the conservation laws of
mass, momentum and energy, one gets the classical Navier-Stokes
equations.

To account for the elastic properties of polymeric fluids, Maxwell combined Newton's law
of viscosity with Hooke's law of elasticity and proposed the following constitutive
relation \cite{Max}
\begin{equation}\label{12}
\epsilon\tau_t + \tau = - \nu\big[\nabla v + (\nabla v)^T -\frac{2}{3}\nabla \cdot v I\big] - \kappa\nabla\cdot v I .
\end{equation}
Here $\epsilon$ is the ratio of the viscosity and the elastic modulus. A Maxwell fluid is
that obeying the constitutive relation (\ref{12}). This relation reflects that the stress
tensor responds to the fluid motion in a delayed, instead of instant, fashion. It has
motivated many more realistic and nonlinear constitutive relations, including the
well-known upper-convected Maxwell (UCM) and Oldroyd-B models \cite{DG}.

In this paper, we revise Maxwell's constitutive relation (\ref{12}), combine the
conservation laws and formulate the following partial differential equations for
compressible viscoelastic fluid flows:
\begin{equation}\label{13}
\begin{array}{rl}
\partial_t\rho + \nabla\cdot(\rho v)& =0,\\[4mm]
\partial_t(\rho v) + \nabla\cdot(\rho v\otimes v + pI)
+\frac{1}{\epsilon_1}\nabla\cdot\tau_1 + \frac{1}{\epsilon_2}\nabla\tau_2 & =0, \\[4mm]
\partial_t\tau_1 + \frac{1}{\epsilon_1}\big[\nabla v
+ (\nabla v)^T -\frac{2}{3}\nabla \cdot v I\big] & =-\frac{\tau_1}{\nu\epsilon_1^2}, \\[4mm]
\partial_t\tau_2 + \frac{1}{\epsilon_2}\nabla\cdot v & = -\frac{\tau_2}{\kappa\epsilon_2^2}.
\end{array}
\end{equation}
Here $\rho$ is the density of the fluid, $\otimes$ denotes the
tensorial product, $p=p(\rho)$ is the hydrostatic pressure, $\tau_1$
is a
tensor of order two, $\epsilon_1$ and $\epsilon_2$ are two positive
parameters, and $\tau_2$ is a scalar. This is a system of
first-order partial differential equations, with domain
$$
G: = \{(\rho, \rho v, \tau_1, \tau_2): \rho>0\}.
$$
In (\ref{13}) there are $14$ equations (for three-dimensional
problems). Note that the $[\cdots]$-term in the $\tau_1$-equation is
symmetric and traceless. It is easy to see that $\tau_1$ is
symmetric and traceless if it is so initially. When $\tau_1$ is
symmetric and traceless, the number of independent equations in
(\ref{13}) reduces to $n=10$. Throughout this paper, we assume that
$\tau_1$ is symmetric and traceless.

We will show that the first-order system (\ref{13}) satisfies the
entropy dissipation condition proposed in \cite{Y3}. This
particularly implies that the system is symmetrizable hyperbolic.
Moreover, we will show that the revised Maxwell's constitutive
relations (the $\tau_1$-, $\tau_2$-equations) in (\ref{13}) are
compatible with Newton's law of viscosity (\ref{11}) for small
$\epsilon_1$ and $\epsilon_2$. To see this, we rewrite the two
$\tau$-equations in (\ref{13}) as follows and iterate them once to
obtain
$$
\begin{array}{rl}
\tau_1 = & -\epsilon_1\nu\big[\nabla v + (\nabla v)^T
-\frac{2}{3}\nabla \cdot v I\big] -
\epsilon_1^2\nu\partial_t\tau_1, \\[4mm]
= & -\epsilon_1\nu\big[\nabla v + (\nabla v)^T -\frac{2}{3}\nabla \cdot v I\big]
+ O(\epsilon_1^3),\\[4mm]
\tau_2 = & -\epsilon_2\kappa\nabla\cdot v - \epsilon_2^2\kappa\partial_t\tau_2, \\[4mm]
= & -\epsilon_2\kappa\nabla\cdot v + O(\epsilon_2^3).
\end{array}
$$
Substituting the truncations into the momentum equation in
(\ref{13}), we obtain the classical isentropic Navier-Stokes
equations. In this sense, Newton's law (\ref{11}) is recovered.

A major part of this paper is devoted to a rigorous justification of
the compatibility above. To do this, we employ the
convergence-stability principle \cite{Y2, BY} for initial-value
problems of symmetrizable hyperbolic systems and prove that, as
$\epsilon_1$ and $\epsilon_2$ go to zero, smooth solutions to the
first-order system
exist in
the time interval where the isentropic Navier-Stokes equations have
smooth solutions and converge to the latter. Namely, we show that
the first-order system (\ref{13}) is a diffusive relaxation
approximation to the isentropic Navier-Stokes equations.



Let us remark that, despite being quite similar, the present problem
is very different from those studied in \cite{LY, Y4}. In fact, when
writing (\ref{13}) in its quasilinear form, the coefficients of
$\frac{1}{\epsilon}$ in the left-hand side depend on $\rho$ and
$\rho v$. Therefore, our problem does not possess the parabolic
structure required in \cite{LY}. It also differs from that in
\cite{Y1}, for $\rho$ and $\rho v$ are not dissipative quantities.
Because of these, our analysis contains some innovative treatments
relying on the specific structure of (\ref{13}).

We end this introduction by mentioning some other related works
known to the author. Studying the motion of complex fluids involves
many challenging and interesting partial differential equations
\cite{RHN} and has attracted much attention in recent years (see
\cite{Mas} and references cited therein). Most mathematical
literature are concerned with well-posedness of incompressible flows
governed by partial differential equations with upper-convected
derivatives included. It seems that there are very few results on
the Newtonian limit of non-Newtonian fluid flows. The only one known
to this author is \cite{MT}, which was concerned with incompressible
viscoelastic fluid flows of Oldroyd type for solutions in the Besov
spaces.

The paper is organized as follows. In Section 2 we show that the
first-order system (\ref{13}) satisfies the entropy dissipation
condition proposed in \cite{Y3}. Section 3 is devoted to a precise
statement of our compatibility result. A key error estimate is
derived in Section 4.

\section{Entropy dissipation structure}
\setcounter{equation}{0}

In this section, we show that the first-order system (\ref{13}) satisfies the entropy
dissipation condition proposed in \cite{Y3}. To this purpose, we define
$$
\Phi(\rho) = \rho\int_1^\rho\frac{p(z)}{z^2}dz
$$
and compute, for smooth solutions to (\ref{13}),
\begin{equation}\label{21}
\begin{array}{rl}
& \partial_t\big(4\Phi(\rho) + 2\rho|v|^2 + 2\tau_2^2 + |\tau_1|^2\big) \\[4mm]
+ & \nabla\cdot\big(4\Phi(\rho)v + 2\rho|v|^2v + 4pv +
4\frac{\tau_2v}{\epsilon_2} + 4\frac{\tau_1v}{\epsilon_1} \big) =
-\frac{4\tau_2^2}{\kappa\epsilon_2^2}
-\frac{2|\tau_1|^2}{\nu\epsilon_1^2}  .
\end{array}
\end{equation}
Here $|\tau_1|^2$ is the trace of the matrix $\tau_1^T\tau_1$ and we
have used the fact that $\tau_1$ is a symmetric and traceless
tensor.
It is easy to verify that
$$
\eta =\eta(U) := 4\Phi(\rho) + 2\rho|v|^2 + 2 \tau_2^2 + |\tau_1|^2
$$
is a strictly convex function of $U := (\rho, \rho v, \tau_1,
\tau_2)^T$, provided that the pressure $p=p(\rho)$ is strictly
increasing with respect to $\rho>0$. Thus, the first-order system
(\ref{13}) fulfills the entropy dissipation condition in \cite{Y3}.
This particularly implies that the system is symmetrizable
hyperbolic.

Set
$$
w = (\rho, \rho v)^T \quad \mbox{and}\quad z = (\tau_1, \tau_2)^T.
$$
We may rewrite the first-order system (\ref{13}) (with $\epsilon_1 =
\epsilon_2 \equiv \epsilon$ for simplicity) as
\begin{equation}\label{22}
\begin{array}{rl}
\partial_tw + \sum_jf_j(w)_{x_j} + \frac{1}{\epsilon}\sum_jC_jz_{x_j} & = 0,\\[4mm]
\partial_tz + \frac{1}{\epsilon}\sum_jg_j(w)_{x_j} & = -\frac{1}{\epsilon^2}Sw,
\end{array}
\end{equation}
where $C_j$ is a constant matrix and $S = \mbox{diag}(\nu^{-1}I_9,
{\kappa}^{-1})$. Moreover, from the form of $\eta=\eta(U)$ and the
strict convexity it follows that $\eta_{ww}(w, z)$ is a symmetric
positive-definite matrix, $\eta_{zz}(w, z)$ is a constant diagonal
(and positive-definite) matrix, $\eta_{ww}(w, z)f_{jw}(w)$ is
symmetric, and
\begin{equation}\label{23}
\eta_{ww}(w, z)C_j=g_{jw}(w)^T\eta_{zz}(w, z)^T.
\end{equation}
The last two statements are based on (\ref{21}).

\section{Compatibility theorem}
\setcounter{equation}{0}

This section is devoted to a precise statement of our compatibility
result. For the sake of simplicity, we assume that $\epsilon_1 =
\epsilon_2 \equiv \epsilon$ in what follows.

Let $\rho=\rho(x, t)$ and $v=v(x, t)$ be the density and velocity of
the Newtonian fluid. Then they obey the conservation laws of mass
and momentum
\begin{equation*}
\begin{array}{rl}
\partial_t\rho + \nabla\cdot(\rho v)& =0,\\[4mm]
\partial_t(\rho v) + \nabla\cdot(\rho v\otimes v + p(\rho)I) +
\nabla\cdot\tau & =0
\end{array}
\end{equation*}
together with Newton's law of viscosity (\ref{11})
$$
\tau = - \nu\Big[\nabla v + (\nabla v)^T -\frac{2}{3}\nabla \cdot v
I\Big] - \kappa\nabla\cdot v I.
$$
Namely, they solve the isentropic Navier-Stokes equations.

Define
$$
\rho_\epsilon = \rho, \quad v_\epsilon = v, \quad \tau_{1\epsilon} =
- \epsilon\nu\Big[\nabla v + (\nabla v)^T -\frac{2}{3}\nabla \cdot v
I\Big], \quad \tau_{2\epsilon} = - \epsilon\kappa\nabla\cdot v.
$$
We have
\begin{equation}\label{31}
\begin{array}{rl}
\partial_t\rho_\epsilon + \nabla\cdot(\rho_\epsilon v_\epsilon)& =0,\\[4mm]
\partial_t(\rho_\epsilon v_\epsilon) + \nabla\cdot(\rho_\epsilon v_\epsilon\otimes v_\epsilon + p(\rho_\epsilon)I) +\frac{1}{\epsilon}\nabla\cdot\tau_{1\epsilon} + \frac{1}{\epsilon}\nabla\tau_{2\epsilon} & =0, \\[4mm]
\partial_t\tau_{1\epsilon} + \frac{1}{\epsilon}\Big[\nabla v_\epsilon +
(\nabla v_\epsilon)^T -\frac{2}{3}\nabla \cdot v_\epsilon I\Big] &
=-\frac{\tau_1}{\nu\epsilon^2} + \partial_t\tau_{1\epsilon}, \\[4mm]
\partial_t\tau_{2\epsilon} + \frac{1}{\epsilon}\nabla\cdot v_\epsilon I&
= -\frac{\tau_{2\epsilon}}{\kappa\epsilon^2} +
\partial_t\tau_{2\epsilon}.
\end{array}
\end{equation}

Our compatibility result can be stated as

\begin{thm}\label{main}
Suppose the pressure function $p=p(\rho)$ is strictly increasing
with respect to $\rho>0$, the density $\rho$ and velocity $v$ of the
Newtonian fluid are continuous and bounded in $(x,
t)\in\Omega\times[0, T_*]$ with $T_*<\infty$, and satisfy
$\inf\limits_{x, t} \rho(x, t)>0$ and
$$
\nabla\rho\in C([0, T_*], H^s(\Omega)), \qquad v\in C^1([0, T_1],
H^{s+1}(\Omega))
$$
with integer $s>2$. Then there exist positive numbers
$\epsilon_0=\epsilon_0(T_*)$ and $K=K(T_*)$ such that for
$\epsilon\leq\epsilon_0$ the first-order system (\ref{13}) with
initial data $(\rho, \rho v,\tau_{1\epsilon}, \tau_{2\epsilon})|_{t
= 0}$ has a unique classical solution $(\rho^\epsilon, \rho^\epsilon
v^\epsilon,\tau_1^\epsilon, \tau_2^\epsilon)$ satisfying
$$
(\rho^\epsilon - \rho, \rho^\epsilon v^\epsilon,\tau_1^\epsilon,
\tau_2^\epsilon)\in C([0, T_*], H^s(\Omega))
$$
and
\begin{equation}\label{32}
\sup_{t\in[0, T_*]}\big\|\big[(\rho^\epsilon, \rho^\epsilon
v^\epsilon,\tau_1^\epsilon, \tau_2^\epsilon) - (\rho, \rho v,
\tau_{1\epsilon}, \tau_{2\epsilon})\big](\cdot, t)\big\|_s\leq
K(T_*) \epsilon^2.
\end{equation}
\end{thm}
\noindent Here $\Omega=\mathbb{R}^3$ or the three-dimensional torus
$[0, 1]^3$, and we are using the standard notation for Sobolev
spaces, defined in \cite{BY, LY, Ma, Y1, Y2, Y3, Y4}.

An obvious corollary of this theorem is
\begin{cor}
If $\rho$ and $v$ possess the properties assumed in Theorem
\ref{main} globally in time, then the time interval where
$(\rho^\epsilon, \rho^\epsilon v^\epsilon,\tau_1^\epsilon,
\tau_2^\epsilon)$ exists goes to infinity as $\epsilon$ tends to
zero.
\end{cor}

To see the existence claim in Theorem \ref{main}, we recall from the
previous section that the first-order system (\ref{13}) is
symmetrizable hyperbolic. Thus, the local-in-time existence theory
of regular solutions to initial-value problems of symmetrizable
hyperbolic systems applies \cite{Ma}. Fix $\epsilon>0$. According to
the local-in-time existence theory, there is a time interval $[0,T]$
such that (\ref{13}) with initial data $(\rho, \rho
v,\tau_{1\epsilon}, \tau_{2\epsilon})|_{t = 0}$ has a unique
solution $(\rho^\epsilon, \rho^\epsilon v^\epsilon,\tau_1^\epsilon,
\tau_2^\epsilon)$ satisfying
\begin{eqnarray*}
(\rho^\epsilon - \rho, \rho^\epsilon v^\epsilon,\tau_1^\epsilon,
\tau_2^\epsilon)\in C([0,T],H^s(\Omega)).
\end{eqnarray*}
Note that the range $G_1$ of $(\rho, \rho v, \tau_{1\epsilon},
\tau_{2\epsilon})(x, t)$ satisfies $G_1\subset\subset G$ for
$\inf\limits_{x, t} \rho(x, t)>0$. For $G_2\subset G$ satisfying
$G_1\subset\subset G_2$, we define
\begin{eqnarray*}
T^\epsilon=\sup\{T>0: (\rho^\epsilon-\rho, \rho^\epsilon
v^\epsilon,\tau_1^\epsilon, \tau_2^\epsilon)\in
 H^s(\Omega), \quad (\rho^\epsilon, \rho^\epsilon v^\epsilon,
 \tau_1^\epsilon, \tau_2^\epsilon)(x, t)\in G_2\}.
\end{eqnarray*}
Namely, $[0,T^\epsilon)$ is the maximal time interval of
$H^s(\Omega)$-existence. Note that $T^\epsilon=T^\epsilon(G_2)$ may
tend to 0 as $\epsilon$ goes to 0.

In order to show that $T^\epsilon>T_*$, we exploit the
convergence-stability lemma \cite{Y2, BY} and only need to prove the
error estimate in (\ref{32}) for $t\in[0, \min\{T_*, T^\epsilon\})$.
In this time interval, both $(\rho^\epsilon, \rho^\epsilon
v^\epsilon,\tau_1^\epsilon, \tau_2^\epsilon)$ and $(\rho, \rho
v,\tau_{1\epsilon}, \tau_{2\epsilon})$ are well defined, regular
enough and take values in the compact set $G_2$.

\section{Error Estimate}
\setcounter{equation}{0}

The purpose of this section is to derive the error estimate in
(\ref{32}) for $t\in[0, \min\{T_*, T^\epsilon\})$. To do this, we
need some classical calculus inequalities in Sobolev spaces
\cite{HK, Ma}.
\begin{lem}
(i). For $s\geq2$, $H^s=H^s(\Omega)$ is an algebra. Namely, if $f,
g\in H^s$, then $fg\in H^s$ and, for all multi-indices $\alpha$ with
$|\alpha|\leq s$,
\begin{eqnarray*}
\|\partial_x^\alpha(fg)\|\leq C_s\|f\|_s\|g\|_s.
\end{eqnarray*}
\noindent Here $C_s$ is a generic constant depending only on $s$.

(ii). For $s\geq3$, let $f\in H^s$ and $g\in H^{s-1}$. Then for all
multi-indices $\alpha$ with $|\alpha|\leq s$, the commutator
$[\partial_x^\alpha, f]g\equiv
\partial_x^\alpha(fg) - f\partial_x^\alpha g\in L^2(\Omega)$
and
\begin{eqnarray*}
\|[\partial_x^\alpha, f]g\|\leq C_s\|\nabla f\|_{s-1}\|g\|_{s-1}.
\end{eqnarray*}

(iii). Let $f(u)$ be a smooth function of $u$. Then for all multi-indices $\alpha$ with $|\alpha|\geq 1$ we have \begin{eqnarray*}
\|\partial_x^\alpha f(u)\|\leq C(|u|_\infty)\|u\|_{|\alpha|},
\end{eqnarray*}
where $C(|u|_\infty)$ is a constant depending on the maximum norm
$|u|_\infty$ of function $u=u(x)$.
\end{lem}

Now we derive the error estimate. With the notation in Section 2:
$$
w^\epsilon =(\rho^\epsilon, \rho^\epsilon v^\epsilon)^T, \quad
z^\epsilon =(\tau_1^\epsilon, \tau_2^\epsilon)^T, \quad w_\epsilon
=(\rho_\epsilon, \rho_\epsilon v_\epsilon)^T, \quad z_\epsilon
=(\tau_{1\epsilon}, \tau_{2\epsilon})^T.
$$
we set
$$
E_1=w^\epsilon - w_\epsilon \quad \mbox{and}\quad E_2=z^\epsilon - z_\epsilon .
$$
From (\ref{22}) and (\ref{31}) we deduce that
\begin{eqnarray}\label{41}
\begin{array}{rl}
\partial_tE_1 + \sum_j(f_j(w^\epsilon) - f_j(w_\epsilon))_{x_j} + \frac{1}{\epsilon}\sum_jC_jE_{2x_j} & = 0, \\[4mm]
\partial_tE_2 + \frac{1}{\epsilon}\sum_j(g_j(w^\epsilon) - g_j(w_\epsilon))_{x_j} &
= -\frac{1}{\epsilon^2}SE_2 -
\partial_t(\tau_{1\epsilon},\tau_{2\epsilon})^T.
\end{array}
\end{eqnarray}
Let $\alpha$ be a multi-index with $|\alpha|\leq s$. Differentiating
the two sides of the last equations with $\partial_x^\alpha$ and
setting
\begin{eqnarray*}
E_{1\alpha}=\partial_x^\alpha E_1 \quad \mbox{and} \quad E_{2\alpha}=\partial_x^\alpha E_2,
\end{eqnarray*}
we obtain
\begin{equation}\label{42}
\begin{array}{rl}
\partial_tE_{1\alpha} + \sum_jA_j(w^\epsilon)E_{1\alpha x_j}
+ \frac{1}{\epsilon}\sum_jC_jE_{2\alpha x_j} & = f^\alpha, \\[4mm]
\partial_tE_{2\alpha} + \frac{1}{\epsilon}\sum_jB_j(w^\epsilon)E_{1\alpha x_j} &
= -\frac{1}{\epsilon^2}SE_{2\alpha} -
\partial_t\partial_x^\alpha(\tau_{1\epsilon},\tau_{2\epsilon})^T
+ \frac{1}{\epsilon}g^\alpha,
\end{array}
\end{equation}
where $A_j(w)=f_{jw}(w), B_j(w)=g_{jw}(w)$,
\begin{eqnarray}
f^\alpha = & \sum_jA_j(w^\epsilon)E_{1\alpha x_j} -
\sum_j\partial_x^\alpha(f_j(w^\epsilon) - f_j(w_\epsilon))_{x_j}, \notag\\[4mm]
g^\alpha = & \sum_jB_j(w^\epsilon)E_{1\alpha x_j}-
\sum_j\partial_x^\alpha(g_j(w^\epsilon) -
g_j(w_\epsilon))_{x_j}.\notag
\end{eqnarray}

Set $D(w)=\eta_{ww}(w, z)$ and $H=\eta_{zz}(w, z)$. Note that
$D(w)A_j(w)$ is symmetric and the identity (\ref{23}) holds.
Multiplying the first equation in (\ref{42}) with
$E_{1\alpha}^TD(w^\epsilon)$ and the second with $E_{2\alpha}^TH$,
summing up the two and integrating the resultant equality over
$\Omega$ gives
\begin{equation}\label{43}
\begin{array}{rl}
&\frac{d}{dt}\int_{\Omega}\big[E^T_{1\alpha}D(w^\epsilon)E_{1\alpha}
+ E^T_{2\alpha}HE_{2\alpha}\big]dx\\[4mm]
=&-\frac{2}{\epsilon^2}\int_{\Omega}E^T_{2\alpha}HSE_{2\alpha} dx +
2\int_{\Omega}E^T_{1\alpha}D(w^\epsilon)f^\alpha dx
- 2\int_{\Omega}E^T_{2\alpha}H\partial_t\partial_x^\alpha(\tau_{1\epsilon},\tau_{2\epsilon})^T dx \\[4mm]
& + \frac{2}{\epsilon}\int_{\Omega}E^T_{2\alpha}Hg^\alpha dx +
\int_{\Omega}E^T_{1\alpha}\big[\partial_tD(w^\epsilon)
+ \sum_j\partial_{x_j}D(w^\epsilon)A_j(w^\epsilon)]E_{1\alpha}dx\\[4mm]
& + \frac{2}{\epsilon}\sum_j\int_{\Omega}E^T_{1\alpha}\partial_{x_j}D(w^\epsilon)C_jE_{2\alpha} dx\\[4mm]
\leq & -\frac{c}{\epsilon^2}\|E_{2\alpha}\|^2 +
\frac{3c}{4\epsilon^2}\|E_{2\alpha}\|^2
+ C\epsilon^4 + C\|E_{1\alpha}\|^2 \\[4mm]
& + C(\|f^\alpha\|^2 + \|g^\alpha\|^2) + |\partial_tD(w^\epsilon)
+ \sum_j\partial_{x_j}D(w^\epsilon)A_j(w^\epsilon)|_\infty\|E_{1\alpha}\|^2\\[4mm]
& +
C|\sum_j\partial_{x_j}D(w^\epsilon)C_j|_\infty^2\|E_{1\alpha}\|^2.
\end{array}
\end{equation}
Here $c$ and $C$ are both generic positive constants, and we have
used that
$\|\partial_x^\alpha\partial_t(\tau_{1\epsilon},\tau_{2\epsilon})\|\leq
C \epsilon$.

Next we analyze the terms in the last two lines. For $\|f^\alpha\|$,
we use Lemma \ref{41} and the boundedess of $\|w_{\epsilon x_j}\|_s$
to estimate as follows.
\begin{equation}\label{44}
\begin{array}{rl}
\|f^\alpha\| & = \|\sum_j[A_j(w^\epsilon)E_{1\alpha x_j}
- \partial_x^\alpha(f_j(w^\epsilon) - f_j(w_\epsilon))_{x_j}]\|\\[4mm]
&\leq \sum_j\|A_j(w^\epsilon)E_{1\alpha x_j}
- \partial_x^\alpha\big(A_j(w^\epsilon)w^\epsilon_{x_j}
- A_j(w_\epsilon)w_{\epsilon x_j}\big)\|\\[4mm]
&\leq \sum_j\|A_j(w^\epsilon)E_{1\alpha x_j} -
\partial_x^\alpha\big(A_j(w^\epsilon)E_{1x_j}
+ (A_j(w^\epsilon) - A_j(w_\epsilon))w_{\epsilon x_j}\big)\|\\[4mm]
&\leq \sum_j\|[A_j(w^\epsilon), \partial_\alpha]E_{1x_j}\|
+ \sum_j\|\partial_x^\alpha\big((A_j(w^\epsilon) - A_j(w_\epsilon))w_{\epsilon x_j}\big)\|\\[4mm]
&\leq C_s\sum_j\|\nabla A_j(w^\epsilon)\|_{s-1}\|E_{1x_j}\|_{s-1}
+ C_s\sum_j\|A_j(w^\epsilon) - A_j(w_\epsilon)\|_s\|w_{\epsilon x_j}\|_s\\[4mm]
&\leq C_s\sum_j\|\nabla A_j(w_\epsilon)\|_{s-1}\|E_1\|_s + C_s\sum_j\|\nabla A_j(w^\epsilon)-\nabla A_j(w_\epsilon)\|_{s-1}\|E_1\|_s \\[4mm]
& \qquad + C_s\sum_j\|A_j(w^\epsilon) - A_j(w_\epsilon)\|_s\\[4mm]
&\leq C_s\|E_1\|_s + C_s(1 + \|E_1\|_s)\sum_j\|A_j(w^\epsilon) - A_j(w_\epsilon)\|_s\\[4mm]
&\leq C_s(1 + \|E_1\|_s)\|E_1\|_s.
\end{array}
\end{equation}
Similarly, we have
\begin{equation}\label{45}
\|g^\alpha\|\leq C_s(1 + \|E_1\|_s)\|E_1\|_s.
\end{equation}
In addition, we have
\begin{equation}\label{46}
\begin{array}{rl}
|\sum_j\partial_{x_j}D(w^\epsilon)C_j| & \leq C\sum_j|w^\epsilon_{x_j}|, \\[4mm]
|\partial_tD(w^\epsilon) + \sum_j\partial_{x_j}D(w^\epsilon)A_j(w^\epsilon)| & \leq C\sum_j|w^\epsilon_{x_j}| + C|w^\epsilon_t|.
\end{array}
\end{equation}
For $|w^\epsilon_t|$ we use the equation and $z_\epsilon=O(\epsilon)$ to get
\begin{equation}\label{47}
\begin{array}{rl}
|w^\epsilon_t| \leq & C\sum_j|w^\epsilon_{x_j}| + C\frac{1}{\epsilon}\sum_j|z^\epsilon_{x_j}|\\[4mm]
\leq & C\sum_j(|w_{\epsilon x_j}| + |w^\epsilon_{x_j} - w_{\epsilon x_j}|) + C\frac{1}{\epsilon}\sum_j(|z_{\epsilon x_j}| + |z^\epsilon_{x_j} - z_{\epsilon x_j}|)\\[4mm]
\leq & C + C\sum_j|E_{1x_j}| + C\frac{1}{\epsilon}\sum_j|E_{2x_j}|\\[4mm]
\leq & C + C\|E_1\|_s + C\frac{1}{\epsilon}\|E_2\|_s.
\end{array}
\end{equation}

Substituting (\ref{44})-(\ref{47}) into (\ref{43}) we arrive at
\begin{equation*}
\begin{array}{rl}
&\frac{d}{dt}\int_{\Omega}\big[E^T_{1\alpha}D(w^\epsilon)E_{1\alpha}+ E^T_{2\alpha}HE_{2\alpha}\big]dx\\[4mm]
\leq & -\frac{c}{4\epsilon^2}\|E_{2\alpha}\|^2 + C\epsilon^4 + C(1+
\|E_1\|_s^2+ \frac{1}{\epsilon}\|E_2\|_s)\|E_1\|_s^2.
\end{array}
\end{equation*}
Summing up this inequality over all $\alpha$ with $|\alpha|\leq s$
gives
$$
\begin{array}{rl}
&\frac{d}{dt}\sum_\alpha\int_{\Omega}\big[E^T_{1\alpha}D(w^\epsilon)E_{1\alpha}+ E^T_{2\alpha}HE_{2\alpha}\big]dx\\[4mm]
\leq & -\frac{c}{4\epsilon^2}\|E_2\|_s^2 + C\epsilon^4 + C(1+ \|E_1\|_s^2+ \frac{1}{\epsilon}\|E_2\|_s)\|E_1\|_s^2\\[4mm]
\leq & -\frac{c}{4\epsilon^2}\|E_2\|_s^2 + C\epsilon^4 + C(1+ \|E_1\|_s^2)\|E_1\|_s^2
+ \frac{c}{8\epsilon^2}\|E_2\|_s^2 + C\|E_1\|_s^4\\[4mm]
\leq & -\frac{c}{8\epsilon^2}\|E_2\|_s^2 + C\epsilon^4 + C(1+ \|E_1\|_s^2)\|E_1\|_s^2.
\end{array}
$$
Integrating the last inequality from 0 to $t\in[0, \min\{T^\epsilon,
T_*\})$ and using the positive definiteness of $D(w^\epsilon)$ and
$H$, we get
\begin{equation}\label{48}
\|E_1\|_s^2+ \|E_2\|^2_s \leq C\epsilon^4 + C\int_0^t(1 +
\|E_1\|_s^2)\|E_1\|_s^2\equiv\phi(t).
\end{equation}
Obviously, $\phi(0)=C\epsilon^4$ and
\begin{equation*}
\phi'(t)= C(1 + \|E_1\|_s^2)\|E_1\|_s^2\leq C\phi(1+ \phi).
\end{equation*}
Applying the nonlinear Gronwall-type inequality in \cite{Y1} to the
last inequality yields
$$
\phi(t)\leq e^{CT_*}
$$
for $t\in[0, \min\{T^\epsilon, T_*\})$ if we choose $\epsilon$ so
small that $\phi(0)=C\epsilon^4\leq e^{-CT_*}$. Finally, we apply
the standard Gronwall inequality to (\ref{48}) to obtain
$$
\|E_1\|_s^2 + \|E_2\|^2_s\leq \phi(t)\leq C\epsilon^4e^{CT_*}.
$$
This completes the proof.

\bigskip\bigskip\bigskip

{\bf{Acknowledgment.}} This work was supported by the Tsinghua
University Initiative Scientific Research Program (20121087902).

\end{document}